\documentclass{article}

\PassOptionsToPackage{numbers}{natbib}
\usepackage[final]{neurips_2022}

\usepackage[utf8]{inputenc} % allow utf-8 input
\usepackage[T1]{fontenc}    % use 8-bit T1 fonts
\usepackage{hyperref}       % hyperlinks
\usepackage{url}            % simple URL typesetting
\usepackage{booktabs}       % professional-quality tables
\usepackage{amsfonts}       % blackboard math symbols
\usepackage{nicefrac}       % compact symbols for 1/2, etc.
\usepackage{microtype}      % microtypography
\usepackage{xcolor}         % colors

\title{Trust Explanations to Do What They Say}

\author{%
  Neil Natarajan \\
  Department of Computer Science\\
  University of Oxford\\
  \texttt{neil.natarajan@new.ox.ac.uk} \\
  \And
  Reuben Binns \\
  Department of Computer Science\\
  University of Oxford\\
  \texttt{reuben.binns@cs.ox.ac.uk} \\
  \And
  Jun Zhao \\
  Department of Computer Science\\
  University of Oxford\\
  \texttt{jun.zhao@cs.ox.ac.uk} \\
  \And
  Nigel Shadbolt \\
  Department of Computer Science\\
  University of Oxford\\
  \texttt{nigel.shadbolt@jesus.ox.ac.uk} \\
}

\begin{document}

\maketitle

\begin{abstract}
  How much are we to trust a decision made by an AI algorithm? Trusting an algorithm without cause may lead to abuse, and mistrusting it may similarly lead to disuse. Trust in an AI is only desirable if it is warranted; thus, calibrating trust is critical to ensuring appropriate use. In the name of calibrating trust appropriately, AI developers should provide contracts specifying use cases in which an algorithm can and cannot be trusted. Automated explanation of AI outputs is often touted as a method by which trust can be built in the algorithm. However, automated explanations arise from algorithms themselves, so trust in these explanations is similarly only desirable if it is warranted. Developers of algorithms explaining AI outputs (xAI algorithms) should provide similar contracts, which should specify use cases in which an explanation can and cannot be trusted.
\end{abstract}

\section{Trust in AI Algorithms}

Increasingly, decisions affecting the lives of lay people are made by AI algorithms. And while these algorithms may be useful, they can also be dangerous. Both unwarranted trust in such an algorithm may lead to wrong and damaging decisions, while unwarranted mistrust in such an algorithm may lead to disuse. Neither outcome is favourable. Hardin draws an important distinction between whether someone or something is trusted and whether that trust is well-placed; i.e. it is worthy of trust \cite{Hardin}. 

It is clear, then, that trust in AI algorithms should be \textit{calibrated}, so that users are led to trust trustworthy AI systems and distrust untrustworthy AI systems. Jacovi et al. propose that trust in AI systems can be understood in terms of a contract between the system and the trustor \cite{Jacovi-et-al}. 

Jacovi et al. define a model of human-AI trust resting on two key properties: the \textit{vulnerability} of the user to the model and the user's ability to \textit{anticipate} the impact of the AI model’s decisions. In the human context, person $A$ trusts person $B$ if and only if $A$ believes that $B$ will act in $A$'s best interest, and $A$ accepts vulnerability to $B$'s actions. In the machine context, we do not always expect the machine to act in our best interests. Instead, user $U$ trusts AI model $M$ if and only if $U$ can anticipate and accepts vulnerability to $M$'s actions \cite{Jacovi-et-al}.

Moreover, trust often does not have a blanket scope; typically, $U$ will trust $M$ regarding some particular actions or range of actions, though a broader trust will include many such actions. In the algorithmic context, this scope is clearly limited – unlike humans, trust in algorithms should never be broad; warranted trust is always scoped to a region in which the algorithm's actions can be anticipated, and in which users might reasonably accept vulnerability to these actions. Generally, this scope is limited to some subsection of the intended use cases of the AI system. Trust placed in an AI system to do something it was not intended to do is often unwarranted; trust placed in an AI system to do something it does not claim to do is always unwarranted. Thus, for an algorithm to be trustworthy in a given scope, that algorithm should demonstrate both that a user can anticipate behaviour in that scope and that the anticipated behaviour is such that users might accept vulnerability to the algorithm. We call this demonstration a \textit{contract}, and call this sort of trust \textit{contractual trust} \cite{Jacovi-et-al}.

Following this framing, the extent to which an algorithm warrants trust is modulated by the extent to which it adheres to its contract. Therefore, when the developers of an algorithm provide a contract regarding the intended use of an algorithm, we can evaluate the trustworthiness of an algorithm by evaluating adherence to the contract.

One method of evaluating adherence to contract comes from a user observing the AI algorithm's reasoning process by way of an explanation or an interpretation. However, unlike human decision-makers, few algorithms are inherently capable of explaining their reasoning. The growing field of Explainable Artificial Intelligence (xAI) aims to develop methods for explaining the reasoning algorithms, often with a broad goal of increasing warranted trust in algorithms. However, though it is clear that these algorithms often increase trust in algorithms, it is not always clear that this trust is warranted, as demonstrated by Jacobs et al. \cite{Jacobs-et-al}. Thus, it seems, there are times where even an explanation of an AI algorithm should be distrusted.

\section{Trust in Explanations of AI Algorithms}

Explanation algorithms help us determine whether to trust AI algorithms, but only if we trust the explanation methods. But when can we trust an explanation algorithm? And what are we trusting it to do? The answer that we are trusting these algorithms to \emph{explain} AI systems is insufficient, because what it means to explain is unspecified. AI explainers can be put to a number of different uses, and different algorithms should be trusted for different uses; contracting to behave appropriately in all of these uses is infeasible (an end-user demands a different explanation than a domain expert), so explanations methods should contract to provide only a particular type of explanation.

Much like AI models themselves, we contend that xAI algorithms should be trusted to uphold specific contracts with respect to the ways in which they are used. For example, a model like \textit{recourse}, developed in Ustun et al., designed to informs end-users of what must be done to change their determination, should not be trusted to report errors in model or to point out the most important features \cite{Ustun-et-al}. Similarly, a model like \textit{Scoped Anchors}, developed in Ribeiro et al., designed to simplify predictions into rule-based approximations, should not be trusted to provide recourse information \cite{Ribeiro-et-al}.

We also contend that xAI methods should be evaluated on whether they can be trusted to do what they say. That is, a good xAI method is one that fulfills its intended use case. Much like trust in AI algorithms, trust in explanations of AI algorithms is contractual; xAI methods should be evaluated in terms of the extent to which they uphold the terms of a contract between the explainer and explainee; and an explainee's trust should be calibrated accordingly. We should not trust an explanation algorithm to do something it has not promised to do. 

The absence of contracts is not a mere conceptual problem; it creates a problematic dialectic and hinders effective critique of xAI methods. To demonstrate this, we consider two particular kinds of AI explanations: SHAP explanations, introduced in Lundberg and Lee, and counterfactual explanations, introduced in Wachter et al. \cite{Lundberg-and-Lee, Wachter-et-al}. Both papers focus on the mathematical properties of the explanation algorithm introduce, but neither makes clear what they contend a good explanation consists in or specifies a circumscribed set of use cases for their methods. We consider two evaluation articles: Kumar et al.'s evaluation of the SHAP method, and Barocas et al.'s evaluation of counterfactual explanations \cite{Kumar-et-al, Barocas-et-al}. Both articles rely on similar notions regarding the purpose of explanations – frameworks that the authors of SHAP and counterfactual explanations do not make clear that they subscribe to. For instance, Barocas et al.'s critique counterfactual explanations on the grounds that they are not useful in providing users with actionable information \cite{Barocas-et-al}. Similarly, Kumar et al. argue that SHAP cannot be used to inform users' actions \cite{Kumar-et-al}.

In this paper we investigate similarities between trust in AI algorithms and trust in explanations of those algorithms (and the outputs they produce). We contend that, like trust in AI algorithms, trust in AI explanation algorithms is composed of an ability to anticipate the algorithm and an acceptance of vulnerability to the algorithm's actions. In both cases, this trust is only desirable if it is warranted. The scope of the trust, in both cases, should be clearly enumerated in a contract, and AI and explanation algorithms alike should be evaluated for trustworthiness within this scope.

\bibliographystyle{plainnat}
\bibliography{main.bib}

\begin{thebibliography}{9}
\providecommand{\natexlab}[1]{#1}
\providecommand{\url}[1]{\texttt{#1}}
\expandafter\ifx\csname urlstyle\endcsname\relax
  \providecommand{\doi}[1]{doi: #1}\else
  \providecommand{\doi}{doi: \begingroup \urlstyle{rm}\Url}\fi

\bibitem[Barocas et~al.(2020)Barocas, Selbst, and Raghavan]{Barocas-et-al}
Solon Barocas, Andrew~D. Selbst, and Manish Raghavan.
\newblock The {Hidden} {Assumptions} {Behind} {Counterfactual} {Explanations}
  and {Principal} {Reasons}.
\newblock \emph{Proceedings of the 2020 Conference on Fairness, Accountability,
  and Transparency}, pages 80--89, January 2020.
\newblock \doi{10.1145/3351095.3372830}.
\newblock URL \url{http://arxiv.org/abs/1912.04930}.
\newblock arXiv: 1912.04930.

\bibitem[Hardin(2002)]{Hardin}
Russell Hardin.
\newblock \emph{Trust and trustworthiness}.
\newblock Russell Sage Foundation, 2002.

\bibitem[Jacobs et~al.(2021)Jacobs, Pradier, McCoy, Perlis, Doshi-Velez, and
  Gajos]{Jacobs-et-al}
Maia Jacobs, Melanie~F. Pradier, Thomas~H. McCoy, Roy~H. Perlis, Finale
  Doshi-Velez, and Krzysztof~Z. Gajos.
\newblock How machine-learning recommendations influence clinician treatment
  selections: the example of antidepressant selection.
\newblock \emph{Translational Psychiatry}, 11\penalty0 (1):\penalty0 1--9,
  February 2021.
\newblock ISSN 2158-3188.
\newblock \doi{10.1038/s41398-021-01224-x}.
\newblock URL \url{https://www.nature.com/articles/s41398-021-01224-x}.
\newblock Bandiera\_abtest: a Cc\_license\_type: cc\_by Cg\_type: Nature
  Research Journals Number: 1 Primary\_atype: Research Publisher: Nature
  Publishing Group Subject\_term: Depression;Scientific community
  Subject\_term\_id: depression;scientific-community.

\bibitem[Jacovi et~al.(2021)Jacovi, Marasović, Miller, and
  Goldberg]{Jacovi-et-al}
Alon Jacovi, Ana Marasović, Tim Miller, and Yoav Goldberg.
\newblock Formalizing {Trust} in {Artificial} {Intelligence}: {Prerequisites},
  {Causes} and {Goals} of {Human} {Trust} in {AI}.
\newblock In \emph{Proceedings of the 2021 {ACM} {Conference} on {Fairness},
  {Accountability}, and {Transparency}}, {FAccT} '21, pages 624--635, New York,
  NY, USA, March 2021. Association for Computing Machinery.
\newblock ISBN 978-1-4503-8309-7.
\newblock \doi{10.1145/3442188.3445923}.
\newblock URL \url{https://doi.org/10.1145/3442188.3445923}.

\bibitem[Kumar et~al.(2020)Kumar, Venkatasubramanian, Scheidegger, and
  Friedler]{Kumar-et-al}
I.~Elizabeth Kumar, Suresh Venkatasubramanian, Carlos Scheidegger, and Sorelle
  Friedler.
\newblock Problems with {Shapley}-value-based explanations as feature
  importance measures.
\newblock \emph{arXiv:2002.11097 [cs, stat]}, June 2020.
\newblock URL \url{http://arxiv.org/abs/2002.11097}.
\newblock arXiv: 2002.11097.

\bibitem[Lundberg and Lee(2017)]{Lundberg-and-Lee}
Scott Lundberg and Su-In Lee.
\newblock A {Unified} {Approach} to {Interpreting} {Model} {Predictions}.
\newblock \emph{arXiv:1705.07874 [cs, stat]}, November 2017.
\newblock URL \url{http://arxiv.org/abs/1705.07874}.
\newblock arXiv: 1705.07874.

\bibitem[Ribeiro et~al.(2018)Ribeiro, Singh, and Guestrin]{Ribeiro-et-al}
Marco~Tulio Ribeiro, Sameer Singh, and Carlos Guestrin.
\newblock Anchors: {High}-{Precision} {Model}-{Agnostic} {Explanations}.
\newblock In \emph{{AAAI}}, 2018.

\bibitem[Ustun et~al.(2019)Ustun, Spangher, and Liu]{Ustun-et-al}
Berk Ustun, Alexander Spangher, and Yang Liu.
\newblock Actionable {Recourse} in {Linear} {Classification}.
\newblock \emph{Proceedings of the Conference on Fairness, Accountability, and
  Transparency}, pages 10--19, January 2019.
\newblock \doi{10.1145/3287560.3287566}.
\newblock URL \url{http://arxiv.org/abs/1809.06514}.
\newblock arXiv: 1809.06514.

\bibitem[Wachter et~al.(2017)Wachter, Mittelstadt, and Russell]{Wachter-et-al}
Sandra Wachter, Brent~D. Mittelstadt, and Chris Russell.
\newblock Counterfactual {Explanations} without {Opening} the {Black} {Box}:
  {Automated} {Decisions} and the {GDPR}.
\newblock \emph{CoRR}, abs/1711.00399, 2017.
\newblock URL \url{http://arxiv.org/abs/1711.00399}.
\newblock \_eprint: 1711.00399.

\end{thebibliography}

\end{document}